\documentclass[]{IEEEtran}
\usepackage{subfigure}
\usepackage{graphicx}
\usepackage{array}
\usepackage{mdwmath}
\usepackage{mdwtab}
\usepackage{eqparbox}
\usepackage{algorithmic}
\usepackage[cmex10]{amsmath}
\usepackage{setspace}
\usepackage{algorithm,caption}
\usepackage{empheq}
\usepackage{cases}
\usepackage{soul}
\usepackage{color}
\usepackage{cite}
\usepackage{amssymb}
\usepackage{enumerate}
\usepackage[neverdecrease]{paralist}
\usepackage{multicol}
\usepackage{stfloats}
\usepackage{flushend}
\usepackage{cuted}
\usepackage{pifont}
\usepackage{array}\usepackage{float}\usepackage{multirow}

\makeatletter
\renewcommand{\maketag@@@}[1]{\hbox{\m@th\normalsize\normalfont#1}}%
\makeatother

\begin{document}
\setlength{\textfloatsep}{4pt}
\title{Performance Analysis for Covert Communications Under Faster-than-Nyquist Signaling}

\author{Yuan Li, Yuchen Zhang, Wanyu Xiang, Jianquan Wang, Sa Xiao, Liang Chang and Wanbin Tang
\thanks{Yuan Li, Yuchen Zhang, Wanyu Xiang, Jianquan Wang, Sa Xiao, and Wanbin Tang are with the National Key Laboratory of Science and Technology on Communications, University of Electronic Science and Technology of China, Chengdu, 611731 China (e-mail: yuanli@std.uestc.edu.cn, yc\_zhang@std.uestc.edu.cn, lynn98715@163.com, wjq2002wjq@126.com, xiaosajordan23@163.com, wbtang@uestc.edu.cn).

Liang Chang is with Beijing ATV Technology Co.,Ltd., Beijing, China (e-mail: changliang@aotev.com).
}
}

\maketitle

\begin{abstract}
In this letter, we analyze the performance of covert communications under \emph{faster-than-Nyquist} (FTN) signaling in the Rayleigh block fading channel. Both Bayesian criterion- and \emph{Kullback-Leibler} (KL) divergence-based covertness constraints are considered. Especially, for KL divergence-based one, we prove that both the maximum transmit power and covert rate under FTN signaling are higher than those under Nyquist signaling. Numerical results coincide with our analysis and validate the advantages of FTN signaling to realize covert data transmission.

\end{abstract}


\IEEEpeerreviewmaketitle
\section{Introduction}
\IEEEPARstart{T}HANKS to the higher symbol rate at the transmitter side, \emph{faster-than-Nyquist} (FTN) signaling has been regarded as a promising technology to improve the \emph{spectrum efficiency} (SE) in the future wireless communication systems \cite{ftn,CC_FTN,pre2}. However, the higher symbol rate can also incur severe \emph{inter-symbol-interference} (ISI), which may degrade the transmit rate if has not been carefully dealt with. Therefore, the ISI cancellation schemes to improve the SE of FTN signaling have been extensively studied \cite{pre2,SVDFTN,CDFTN,SS2021,filterhop,VSDFTN}. 

On the other hand, besides degrading the transmit rate of the legitimate data transmission, the incurred ISI can also aggravate the detection or decoding of illegal eavesdroppers. Therefore, some existing work has exploited the ISI incurred by FTN signaling to enhance the security of the data transmission \cite{SS2021,filterhop,VSDFTN}. In \cite{SS2021}, the eigenvalue decomposition based scheme had been investigated to improve the secrecy rate of FTN signaling in quasi-static fading channel. In \cite{filterhop} and \cite{VSDFTN}, the time-varied coefficients of shaped filter and symbol duration were introduced to incur more severe interference to ensure secure transmission, respectively. However, the aforementioned work all focuses on exploiting FTN signaling to avoid the information from being decoded by illegal eavesdroppers, i.e., physical layer security. 
In contrast, covert communications, which aim to prevent the communication between the transmitter Alice and the receiver Bob, from being detected by the eavesdropper Willie, can provide stronger security \cite{Bash, S_Lee, Bash1, Bloch, YSH}. For instance, in military wireless communications, preventing communications from being detected has a higher priority. Also, Alice can transmit no more than $\mathcal{O}(\sqrt{n})$ bits to Bob covertly and reliably for $n$ channel uses, i.e., the square root law \cite{Bash}, has been proved in covert communications under \emph{additive white Gaussian noise} (AWGN) channel. To the best of our knowledge, the covert issue has not been studied under FTN signaling. Therefore, the covert performance of FTN signaling is worth further exploring.

In this letter, we analyze the performance of covert communications under FTN signaling in the Rayleigh block fading channel. We consider both the Bayesian criterion- and \emph{Kullback-Leibler} (KL) divergence-based covertness constraints, and theoretically prove the transmit power and covert rate under FTN signaling with the latter constraint are higher than those under Nyquist signaling. Numerical results coincide with our analysis and show that the maximum transmit power and covert rate can be increased via FTN signaling and decrease with the value of the acceleration factors. 

\section{System Model}
\begin{figure}[!htbp]
\centering
\includegraphics[width=3.0 in]{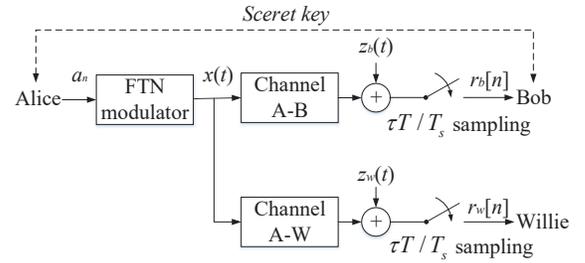}
\caption{The system model of covert communications under FTN siganling.}\label{SM}
\end{figure}
In this paper, we consider a covert communication system under FTN signaling as shown in Fig. \ref{SM}, where a legitimate transmitter Alice transmits signals to its receiver Bob, while an eavesdropper Willie wants to detect the existence of the data transmission. In addition, there is a secret key shared between Alice and Bob to ensure their communication \cite{Bash1, Bloch}. For instance, the secret key can help Bob synchronize data or choose codebook. We assume that channel A-B and A-W follow the Rayleigh block fading.
The transmitted signals of Alice can be represented by
\begin{equation}
x(t)=\sum_{n=-\infty}^{\infty}{a_nh(t-n\tau T)},
\end{equation}
where $a_n \sim \mathcal{CN}(0,\sigma_a^2)$ is the transmitted symbol. This assumption is widely adopted in literature \cite{CC_FTN, pre2, SVDFTN} to obtain the capacity of FTN signaling. In addition, we use this assumption to obtain the upper bound of the performance analysis of using FTN signaling in covert communications, which provides a theoretical guidance for practical cases. $\tau \in (0,1)$ is the acceleration factor, $h(t)$ is the corresponding \emph{root raised cosine} (RRC) shaped filter with unit energy and roll-off factor $\alpha$. $T$ is the orthogonal interval of $h(t)$. Accordingly, the transmit power $P$ is
\begin{equation}
P=\frac{\sigma_a^2}{\tau T}.
\end{equation}

In this letter, in order to provide more stringent covertness requirement for the legitimate data transmission, $h(t)$ and $\tau$ are assumed to be perfectly known by Willie. After the process of matched filter, the received signal can be represented by
\begin{equation}
r_w(t) = \sum_{n=-\infty}^{\infty}{h_{aw}a_ng(t-n\tau T)}+\eta_w(t),
\end{equation}
where $h_{aw}\sim \mathcal{CN}(0,1)$ is the fading coefficients of channel A-W, $g(t)=h(t)*h^*(-t)$, $\eta_w(t)=z_w(t)*h^*(-t)$, and $z_w(t)$ is the AWGN at Willie's side.

The hypothesis test at Willie's side can be represented by
\begin{equation}
\left\{
\begin{aligned}
\mathcal{H}_0:~ &r_w[n]=\eta_w[n] \\
\mathcal{H}_1:~ &r_w[n]=\sum_{l=-\infty}^{\infty}h_{aw}a[n-l]g[l]+\eta_w[n]
\end{aligned}
\right.
\label{HT_o}
\end{equation}
according to the Ungerboeck observation model \cite{Ungerboeck}. Since $r_w[n]$ is correlated to each other, we apply $\textbf{r}_w\in\mathbb{C}^{N\times 1}$ as the observation value, where $N$ is the received symbol block length. Thus, (\ref{HT_o}) can be rewritten as
\begin{equation}
\left\{
\begin{aligned}
\mathcal{H}_0:~ &\textbf{r}_w=\boldsymbol{\eta}_w \\
\mathcal{H}_1:~ &\textbf{r}_w=h_{aw}\textbf{G}\textbf{a}+\boldsymbol{\eta}_w,
\end{aligned}
\right.
\label{HT}
\end{equation}
where $\boldsymbol{\eta}_w\sim \mathcal{CN}(\textbf{0},\sigma_w^2\textbf{G}$), $\textbf{G}\in\mathbb{R}^{N\times N}$ is the corresponding positive definite ISI matrix \cite{SVDFTN, CDFTN}, and $E[\textbf{aa}^H]=\sigma_a^2\textbf{I}$.

\section{Covertness Constraints Analysis}
In this section, we analyze the covertness constraints under Bayesian criterion and KL divergence with FTN signaling.
\subsection{Covertness Constraint Analysis under Bayesian Criterion}
The likelihood function of $\textbf{r}_w$ can be represented by 
\begin{equation}
L(\textbf{r}_w)=\frac{p_1(\textbf{r}_w)}{p_0(\textbf{r}_w)},
\label{LF}
\end{equation}
where $p_0(\textbf{r}_w)=p\left( \textbf{r}_w|\mathcal{H}_1\right)$ and $p_1(\textbf{r}_w)=p\left( \textbf{r}_w|\mathcal{H}_0\right)$ are the \emph{probability density function} (PDF) of $\textbf{r}_w$ under $\mathcal{H}_0$ and $\mathcal{H}_1$, respectively.

For Willie, the detection problem is to distinguish $\mathcal{H}_0$ and $\mathcal{H}_1$. Assuming $\mathbb{P}(\mathcal{H}_0)=\mathbb{P}(\mathcal{H}_1)=0.5$, which is widely adopted in detection theory. Then, in the sense of minimizing the error probability, the optimal detector is the likelihood ratio test by the Bayesian criterion \cite{FSSP}, which can be represented as
\begin{equation}
L\left(\textbf{r}_w\right)\mathop{\gtrless}\limits_{\mathcal{D}_0}^{\mathcal{D}_1}1,
\label{DR}
\end{equation}
where $\mathcal{D}_0$ and $\mathcal{D}_1$ are the decision regions of $\mathcal{H}_0$ and $\mathcal{H}_1$, respectively. Then, the false alarm and miss detection probabilities of Willie are $\mathbb{P}_{FA}=\text{Pr}(\mathcal{D}_1|\mathcal{H}_0)$ and $\mathbb{P}_{MD}=\text{Pr}(\mathcal{D}_0|\mathcal{H}_1)$, respectively.

Since $E[\boldsymbol{\eta}_w\boldsymbol{\eta}_w^H]=\sigma_w^2\textbf{G}$ and $E[(h_{aw}\textbf{Ga}+\boldsymbol{\eta}_w)(h_{aw}\textbf{Ga}+\boldsymbol{\eta}_w)^H]=|h_{aw}|^2\sigma_a^2\textbf{G}\textbf{G}^H+\sigma_w^2\textbf{G}$, we have
\begin{equation}
p_0(\textbf{r}_w) = \frac{1}{\pi^N\text{det}(\sigma_w^2\textbf{G})}\text{exp}\lbrace -\textbf{r}_w^H(\sigma_w^2\textbf{G})^{-1}\textbf{r}_w\rbrace,
\label{f0}
\end{equation}
and
\begin{equation}
p_1(\textbf{r}_w) = \frac{\text{exp}\lbrace - \textbf{r}_w^H(|h_{aw}|^2\sigma_a^2\textbf{G}\textbf{G}^H+\sigma_w^2\textbf{G})^{-1}\textbf{r}_w\rbrace}{\pi^N\text{det}(|h_{aw}|^2\sigma_a^2\textbf{G}\textbf{G}^H+\sigma_w^2\textbf{G})}.
\label{f1}
\end{equation}

By substituting (\ref{f0}) and (\ref{f1}) into (\ref{LF}), the logarithm of $L(\textbf{r}_w)$ can be written as
\begin{equation}
l(\textbf{r}_w)=\text{log}\beta_1+\textbf{r}_w^H \left[ \frac{\textbf{G}^{-1}}{\sigma_w^2}-(|h_{aw}|^2\sigma_a^2\textbf{G}\textbf{G}^H+\sigma_w^2\textbf{G})^{-1}\right]\textbf{r}_w,
\label{llr}
\end{equation}
where
\begin{equation}
\beta_1 = \frac{\text{det}( \sigma_w^2\textbf{G})}{\text{det}(|h_{aw}|^2\sigma_a^2\textbf{GG}^H+\sigma_w^2\textbf{G})}.
\label{beta1}
\end{equation}

To be noticed, when $\tau<1/(1+\alpha)$ and $N\to \infty$, \textbf{G} has some asymptotically-zero eigenvalues \cite{pre2}. However, asymptotically-zero is not zero. Therefore, the derivations in this section hold for any feasible $\tau$ theoretically.

As mentioned in \cite{Bash, WJQ, YSH, ZXY}, the covertness constraint under Bayesian criterion is given by
\begin{equation}
\xi^{min}=\mathbb{P}_{FA}+\mathbb{P}_{MD}\geq 1-\epsilon, 
\label{Bayesian_rule}
\end{equation}
where $\epsilon\in[0,1)$ is a given value. 

Furthermore, we have the following theorem for FTN signaling.

\emph{Theorem 1:} Let $\textbf{G}$ = $\textbf{V}^T\boldsymbol{\Lambda}\textbf{V}$, where $\textbf{V}^T\textbf{V}$ = $\textbf{I}$ and $\boldsymbol{\Lambda}$ = \text{diag}$\lbrace$$\lambda_0$, $\lambda_1$, $\cdots$, $\lambda_{N-1}$$\rbrace$. Then, the false alarm and miss detection probabilities of Willie can be represented by
\begin{equation}
\mathbb{P}_{FA}=\frac{1}{2\pi}\int_{\theta}^{\infty}\int_{-\infty}^{\infty} \prod_{n=0}^{N-1}{\frac{1}{1-2j\alpha_n\omega}\text{exp}(-j\omega t)}d\omega dt
\label{PFA}
\end{equation}
and 
\begin{equation}
\mathbb{P}_{MD} =\frac{1}{2\pi}\int^{\theta}_{-\infty}\int_{-\infty}^{\infty} \prod_{n=0}^{N-1}{\frac{1}{1-2j\gamma_n\omega}\text{exp}(-j\omega t)}d\omega dt,
\label{PMD}
\end{equation}
respectively, where $\theta=-\text{log}\beta_1$,
\begin{equation}
\alpha_n = \frac{1}{2}\left( \frac{|h_{aw}|^2\sigma_a^2\lambda_n}{|h_{aw}|^2\sigma_a^2\lambda_n+\sigma_w^2}\right),
\end{equation}
and
\begin{equation}
\gamma_n = \frac{1}{2}\left( \frac{|h_{aw}|^2\sigma_a^2}{\sigma_w^2}\lambda_n\right). 
\end{equation}

\emph{Proof:} Replace $\textbf{G}$ as $\textbf{V}^T\boldsymbol{\Lambda}\textbf{V}$, (\ref{llr}) can be represented by
\begin{equation}
l(\textbf{r}_w) = \text{log}\beta_1+T(\textbf{r}_w),
\end{equation}
where
\begin{equation}
\beta_1 = \prod_{n=0}^{N-1}{\frac{\sigma_w^2}{|h_{aw}|^2\sigma_a^2\lambda_n+\sigma_w^2}}
\end{equation}
and  
\begin{equation}
T(\textbf{r}_w)=\textbf{r}_w^H\textbf{V} \left[ \frac{\boldsymbol{\Lambda}^{-1}}{\sigma_w^2}-(|h_{aw}|^2\sigma_a^2\boldsymbol{\Lambda}^{2}+\sigma_w^2\boldsymbol{\Lambda})^{-1}\right]\textbf{V}^T\textbf{r}_w
\end{equation}
is the test statistic.

Let $\textbf{y}=\textbf{V}^T\textbf{r}_w$, the PDF of $\textbf{y}$ can be represented by 
\begin{equation}
\textbf{y}\sim \left\{
\begin{aligned}
&\mathcal{H}_0:~\mathcal{CN}(0,\sigma_w^2\boldsymbol{\Lambda})\\
&\mathcal{H}_1:~\mathcal{CN}(0,|h_{aw}|^2\sigma_a^2\boldsymbol{\Lambda}^2+\sigma_w^2\boldsymbol{\Lambda}).
\end{aligned}
\right.
\label{y_pdf}
\end{equation}
Then, the logarithm of (\ref{DR}) can be rewritten as
\begin{equation}
T_1(\textbf{y})\mathop{\gtrless}\limits_{\mathcal{D}_0}^{\mathcal{D}_1} \theta,
\end{equation}
where
\begin{align}
T_1(\textbf{y})&=\textbf{y}^H \left[ \frac{\boldsymbol{\Lambda}^{-1}}{\sigma_w^2}-(|h_{aw}|^2\sigma_a^2\boldsymbol{\Lambda}^{2}+\sigma_w^2\boldsymbol{\Lambda})^{-1}\right]\textbf{y}\\
&=\sum_{n=0}^{N-1}\left( \frac{|h_{aw}|^2\sigma_a^2}{\sigma_w^2(|h_{aw}|^2\sigma_a^2\lambda_n+\sigma_w^2)}\right)|y[n]|^2.
\end{align}

Furthermore, the false alarm probability of Willie can be represented by
\begin{equation}
\mathbb{P}_{FA}=\text{Pr}\lbrace T_1(\textbf{y})\geq \theta|\mathcal{H}_0 \rbrace.
\label{pf1}
\end{equation}

Let $v[n]=y[n]/\sqrt{\frac{\sigma_w^2\lambda_n}{2}}$, we have $v[n]\sim \mathcal{CN}(0,2)$. Then (\ref{pf1}) can be rewritten as
\begin{equation}
\mathbb{P}_{FA}=\text{Pr}\lbrace\sum_{n=0}^{N-1}\alpha_n |v[n]|^2\geq \theta|\mathcal{H}_0 \rbrace.
\label{pf2}
\end{equation}

Furthermore, according to Appendix 5A of \cite{FSSP} and \cite{Johnson}, the characteristic function of $T_1(\textbf{y})$ can be represented by
\begin{align}
\phi_{T_1}(\omega) &= E[\text{exp}(j\omega T_1)]\\
&=\prod_{n=0}^{N-1}{\frac{1}{1-2j\alpha_n\omega}}.
\end{align}
Then, the PDF of $T_1(\textbf{y})$ can be represented by
\begin{equation}
p_T(t)=\frac{1}{2\pi}\int_{-\infty}^\infty \phi_T(\omega)\text{exp}(-j\omega t)d\omega, t>0.
\end{equation}
Therefore, (\ref{pf2}) can be rewritten as (\ref{PFA}).

Similarly, let $v[n]=y[n]/\sqrt{\frac{|h_{aw}|^2\sigma_a^2\lambda_n^2+\sigma_w^2\lambda_n}{2}}$, and $\mathbb{P}_{MD}$ can be represented by (\ref{PMD}).$\hfill\blacksquare$

It should be noted that $\mathbb{P}_{FA}$ and $\mathbb{P}_{MD}$ cannot be directly derived from (\ref{PFA}) and (\ref{PMD}), respectively. Therefore, we carry out Monte Carlo simulations to calculate them according to (\ref{pf1}) and $\mathbb{P}_{MD}=\text{Pr}\lbrace T_1(\textbf{y})< \theta|\mathcal{H}_1 \rbrace$, respectively. 
\subsection{Covertness Constraint Analysis under KL divergence}
In this subsection, we first derive the KL divergence-based covertness constraint under FTN signaling, and then compare the maximum transmit powers under FTN and Nyquist signaling.

For the optimal test at Willie, 
\begin{equation}
\xi^{min}= 1-\mathcal{V}_T(p_0,p_1),
\label{OT}
\end{equation}
where $\mathcal{V}_T(p_0,p_1)$ is the total variation between $p_0(\textbf{r}_w)$ and $p_1(\textbf{r}_w)$. To avoid the intractable expressions for $\mathcal{V}_T(p_0,p_1)$, we use KL divergence that is widely adopted in the literature to limit the detection performance at Willie \cite{Bash, YSH}. Specially, according to Pinsker's inequality, we have
\begin{equation}
\mathcal{V}_T(p_0,p_1)\leq \sqrt{\mathcal{D}(p_1||p_0)/2},
\label{Pinsker}
\end{equation}
where 
\begin{equation}
\mathcal{D}(p_1||p_0)=\int_{\textbf{r}_w} p_1(\textbf{r}_w)\text{log}\frac{p_1(\textbf{r}_w)}{p_0(\textbf{r}_w)}d \textbf{r}_w
\label{RE}
\end{equation}
is the KL divergence from  $p_1(\textbf{r}_w)$ to $p_0(\textbf{r}_w)$.
Then, (\ref{Bayesian_rule}) can be obtained by (\ref{OT}) and (\ref{Pinsker}) as
\begin{equation}
\mathcal{D}(p_1||p_0)\leq 2\epsilon^2.
\label{KL_rule}
\end{equation}
To be noticed, the KL divergence-based covertness constraint is stricter than (\ref{Bayesian_rule}), which means it is fully operational in practice \cite{YSH}. Then we have the following theorem.

\emph{Theorem 2:} The KL divergence-based covertness constraint under FTN signaling can be represented by
\begin{align}
\nonumber
&\mathcal{D}(p_1||p_0) = \\
&\sum_{n=0}^{N-1}{\frac{|h_{aw}|^2\sigma_a^2}{\sigma_w^2}\lambda_n-\text{log}\left( \frac{|h_{aw}|^2\sigma_a^2}{\sigma_w^2}\lambda_n+1\right) } \leq 2\epsilon^2,
\label{D2}
\end{align}
where $\lambda_n$ is the eigenvalue mentioned in \emph{Theorem 1}.

\emph{Proof:} By substituting (\ref{f0}) and (\ref{f1}) into (\ref{RE}), we have
\begin{align}
\nonumber
&\mathcal{D}(p_1||p_0) =  \text{log}\beta_1 \\
&+ E\left[\textbf{r}_w^H \left[ \frac{\textbf{G}^{-1}}{\sigma_w^2}-(|h_{aw}|^2\sigma_a^2\textbf{G}\textbf{G}^H+\sigma_w^2\textbf{G})^{-1}\right]\textbf{r}_w\right].
\label{D}
\end{align}

According to \cite{QE}, $E[\textbf{x}^T\textbf{Q}\textbf{x}]=\text{Tr}(\textbf{Q}\boldsymbol{\Sigma}) + \boldsymbol{\mu}^T\textbf{Q}\boldsymbol{\mu}$, where $\boldsymbol{\mu}$ and $\boldsymbol{\Sigma}$ are the mean and covariance of \textbf{x}, respectively. Then (\ref{D}) can be written as
\begin{equation}
\mathcal{D}(p_1||p_0) = \text{log}\beta_1 + \text{Tr}\left( \frac{|h_{aw}|^2\sigma_a^2}{\sigma_w^2}\textbf{G}^H\right),
\end{equation}
and thus we have (\ref{D2}). $\hfill\blacksquare$

\section{Covert Rate Analysis}
According to \cite{CC_FTN,pre2,SVDFTN}, the instantaneous maximum achievable rate of the communication link between Alice and Bob under FTN signaling can be represented by
\begin{equation}
R_{FTN} = \int_{-1/2\tau T}^{1/2\tau T}\text{log}_2\left( 1+\frac{2|h_{ab}|^2P}{N_0}H_{fo}(f)\right) df,
\label{RFTN}
\end{equation}
where $h_{ab}\sim \mathcal{CN}(0,1)$ is the fading coefficient of channel A-B, $H_{fo}(f)=\sum_{k=-\infty}^{\infty}{|H(f-k/(\tau T))|^2}$ and $N_0$ is one-side spectral density of noise, and $H(f)$ is the Fourier transform of $h(t)$.

Following Section III, the instantaneous covert rate under Bayesian criterion- or KL divergence-based covertness constraint can be represented by
\begin{align}
&\max_{P}~R_{FTN} \label{MCR}\\
\nonumber
&\text{s.t.} ~ (\ref{Bayesian_rule})~\text{or}~(\ref{D2}). 
\end{align}

Since $R_{FTN}$ in (\ref{RFTN}) is monotonically increasing with $P$, the optimal $P$ to Problem (\ref{MCR}) should be the maximum transmit power satisfying (\ref{Bayesian_rule}) or (\ref{D2}). For (\ref{Bayesian_rule}), the maximum transmit power can be obtained by solving (\ref{PFA}) and (\ref{PMD}) via MATLAB. For (\ref{D2}), $P_{F}^{max}$ can be obtained from (\ref{D_FTN}). By substituting the maximum transmit power into (\ref{RFTN}), we can obtain the instantaneous maximum covert rate under FTN signaling. Specially, for the KL divergence based case, we have the following theorem.

\emph{Theorem 3:} Let $R_N$ be the instantaneous covert rate of Nyquist signaling under the KL divergence-based covertness constraint. For a given $\epsilon$ satisfying (\ref{KL_rule}), we have $R_{FTN}> R_{N}$.

\emph{Proof:} Assume $P_{F}$ and $P_{N}$ to be the transmit powers of Alice under FTN and Nyquist signaling with the KL divergence-based covertness constraint, respectively. For the same detection time duration at Willie, (\ref{D2}) can be rewritten as
\begin{align}
\nonumber
&\mathcal{D}_1=\\
&\sum_{n=0}^{N-1}{\frac{|h_{aw}|^2\tau TP_{F}\lambda_n}{\sigma_w^2}-\text{log}\left( \frac{|h_{aw}|^2\tau TP_{F}\lambda_n}{\sigma_w^2}+1\right) } \leq 2\epsilon_1^2
\label{D_FTN}
\end{align}
and
\begin{equation}
\mathcal{D}_2=\sum_{n=0}^{N'-1}{\frac{|h_{aw}|^2P_{N} T}{\sigma_w^2}-\text{log}\left( \frac{|h_{aw}|^2P_{N} T}{\sigma_w^2}+1\right) } \leq 2\epsilon_2^2
\label{D_N}
\end{equation}
for FTN and Nyquist signaling, respectively, where $N'=N\tau$. 

Recall $\lambda_n\approx(\tau T)^{-1}H_{fo}(f_n)$ in \cite{pre2}, where 
$f_n=n/N\tau T$. For $N\to\infty$, (\ref{D_FTN}) can be rewritten as
\begin{align}
&\lim_{N\to\infty}\mathcal{D}_1=\lim_{N\to\infty}\sum_{n=0}^{N-1}{\rho P_F H_{fo}(f_n)-\text{log}\left( \rho P_FH_{fo}(f_n)+1\right) } \\
\label{D_FTN2} 
&=\lim_{N\to\infty}N\tau T \int_0^{\frac{1}{\tau T}}{\rho P_FH_{fo}(f)-\text{log}\left( \rho P_FH_{fo}(f)+1\right) df}\\
\label{D_FTN3} 
&=\lim_{N\to\infty}N\tau T\Delta_1 \leq 2\epsilon_1^2,
\end{align}
where $\Delta_1=\rho P_F-\int_0^{\frac{1}{\tau T}}{\text{log}\left( \rho P_FH_{fo}(f)+1\right)df}$ and $\rho=\frac{|h_{aw}|^2}{\sigma_w^2}$. To be noticed, (\ref{D_FTN2}) to (\ref{D_FTN3}) is according to equation (29) in \cite{pre2}. 

Similarly, (\ref{D_N}) with $N\to\infty$ can be rewritten as 
\begin{equation}
\lim_{N\to\infty}\mathcal{D}_2=N\tau T \Delta_2,
\end{equation}
where $\Delta_2 = \int_0^{\frac{1}{T}}{\left( \rho P_NT-\text{log}\left( \rho P_NT+1\right)\right)df}$. Since $\mathcal{D}_1$ and $\mathcal{D}_2$ are both monotonically increasing with respect to the transmit power, $P_F\in (0,P_F^{max}]$ and $P_N\in (0,P_N^{max}]$, respectively, where $P_F^{max}=\mathcal{D}_1^{-1}(\epsilon_1)$ and $P_N^{max}=\mathcal{D}_2^{-1}(\epsilon_2)$. Then, we have
\begin{equation}
\lim_{N\to\infty}\frac{\text{max}\lbrace\mathcal{D}_1\rbrace}{\text{max}\lbrace\mathcal{D}_2\rbrace}=\frac{\Delta_1(P_F^{max},\tau)}{\Delta_2(P_N^{max})}=\frac{\epsilon_1^2}{\epsilon_2^2}.
\end{equation}

It is easy to know that $\Delta_1=\Delta_2$ when $P_F=P_N$ and $\tau=1$. Meanwhile, it is worth noting that $\int_0^{\frac{1}{\tau T}}{\text{log}\left( \rho P_FH_{fo}(f)+1\right)df}$ in $\Delta_1$ is closely related to (\ref{RFTN}), which means $\Delta_1$ increases as $\tau$ increases for $\tau\geq 1/(1+\alpha)$ and stays the same for $\tau< 1/(1+\alpha)$.  

Therefore, when $\epsilon$ is given, i.e., $\epsilon_1=\epsilon_2$ and $\tau<1$, we have $P_{F}^{max}> P_{N}^{max}$.

In addition, for the same transmit power, $R_{FTN}>R_N$ has been proved in \cite{CC_FTN}. Since $R_{FTN}$ and $R_N$ are both monotonically increasing with respect to the transmit power, we have $R_{FTN}>R_N$ for a given $\epsilon$. $\hfill\blacksquare$

Although the superiority of FTN signaling over Nyquist signaling in covert communications is proved for $N\to\infty$, the conclusion still holds for a practical value of $N$, which will be shown in section V.

Correspondingly, from a long-term perspective, the ergodic covert rate can be represented as
\begin{align}
&\max_{P}~E_{|h_{ab}|^2}[R_{FTN}]\label{MCR_fading}\\
\nonumber
&\text{s.t.} ~ E_{|h_{aw}|^2}[\xi^{min}]\geq 1-\epsilon~\text{or}~E_{|h_{aw}|^2}[\mathcal{D}(p_1||p_0)\leq 2\epsilon^2] .
\end{align}

Recall \emph{Theorem 3}, FTN singaling has a higher instantaneous covert rate than the Nyquist signaling for specific $h_{aw}$ and $h_{ab}$. It is straightforward that the ergodic covert rate of FTN signaling is higher than that of the Nyquist signaling.
However, it is troublesome to analysis the advantage of using FTN signaling in covert communications over the Nyquist signaling under the Bayesian criterion-based covertness constraint directly. Therefore, the covert performance of using FTN signaling in the block fading channel will also be evaluated in the next section.
\section{Numerical Results}
In this section, we evaluate the performance of the covert communication link under FTN signaling via the simulation.  
The parameter settings are listed in Table \ref{T}.
\begin{table}[!hbtp]
\renewcommand{\arraystretch}{0.8}
\caption{The parameter settings for the simulation.}
\label{T}
\centering
\begin{tabular}{|c||c|}
\hline
Parameter & Value\\
\hline
Shaped filter type & RRC \\
\hline
$\alpha$ & $0.3$ \\
\hline
$T$ & $1$ second \\
\hline
$\tau$ & $0.5$ to $1$ with step of $0.05$ \\
\hline
$\epsilon$ & $\lbrace 0.01, 0.05\rbrace$ \\
\hline
$N'$ & $\lbrace 1000, 2000, \cdots, 9000, 10000\rbrace$\\
\hline
$h_{ab}$, $h_{aw}$ &  $h_{ab}\sim \mathcal{CN}(0,1)$, $h_{aw}\sim \mathcal{CN}(0,1)$\\
\hline
$\sigma_w^2$ & $1$ W \\
\hline
$N_0$ & 2 W/Hz\\
\hline
\end{tabular}
\end{table}

\begin{figure}[!htbp]
\centering
\includegraphics[width=3.0 in]{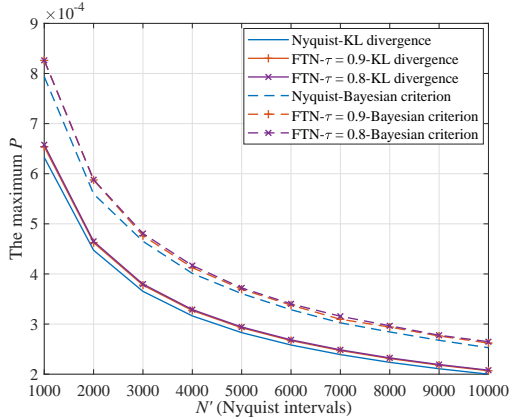}
\caption{The maximum $P$ versus Nyquist intervals for $\epsilon=0.01$.}\label{P_N_new}
\end{figure}
\begin{figure}[!htbp]
\centering
\includegraphics[width=3.0 in]{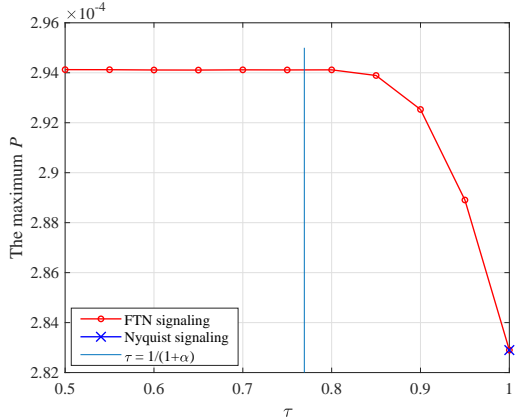}
\caption{The maximum $P$ versus $\tau$ for $N'=5000$ and $\epsilon=0.01$.}\label{P_tau}
\end{figure}

Figure \ref{P_N_new} shows $P$ versus $N'$ for $\epsilon=0.01$ under different covertness constraints in the AWGN channel with the same detection time of both schemes, where $N'$ is the Nyquist intervals, i.e., $N=N'$ and $N = \lceil N'/\tau \rceil$ for Nyquist and FTN signaling, respectively. Although \emph{Theorem 3} is proved for $N\to\infty$, the advantages of FTN signaling over Nyquist signaling still exist for large enough $N$. However, the gain of maximum $P$ of FTN increases slowly as the decrease of $\tau$. Therefore, we also plot the maximum $P$ versus $\tau$ for $N' = 5000$ and $\epsilon=0.01$ in Fig. \ref{P_tau}, where $\tau=1$ corresponds to the Nyquist signaling. It can be observed that the maximum $P$ increases as $\tau$ decreases for $\tau\geq 1/(1+\alpha)=1/1.3$ and stays the same for $\tau<1/1.3$, which is coincide with the proof of \emph{Theorem 3}.

\begin{figure}[!htbp]
\centering
\includegraphics[width=3.0 in]{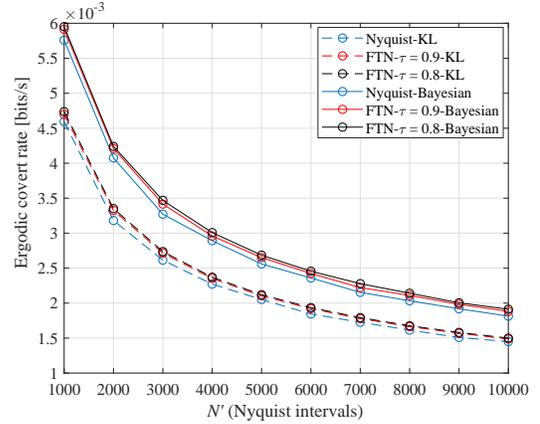}
\caption{The ergodic covert rate versus $N'$ for $\epsilon=0.05$ in the Rayleigh block fading channel.}\label{C_N_fading}
\end{figure}

Figure \ref{C_N_fading} shows the ergodic covert rate under FTN and Nyquist signaling versus $N'$ for $\epsilon=0.05$ in the Rayleigh block fading channel. As we can see, the ergodic covert rate under FTN signaling is higher than that under the Nyquist signaling for both covertness constraints. 

\vspace{4ex}
\section{Conclusion}
In this letter, we have analyzed the performance of covert communications under FTN signaling in the Rayleigh block fading channel. 
Both Bayesian criterion- and KL divergence-based covertness constraints have been considered. 
Especially, for KL divergence-based one, we have proved that the maximum transmit power and covert rate were higher using FTN signaling than those under Nyquist signaling.
Numerical results have coincided with our analysis and validated the advantages of FTN signaling to realize covert data transmission.


\end{document}